# Effective immunity and second waves: a dynamic causal modelling study

Karl J. Friston[1], Thomas Parr[1], Peter Zeidman[1], Adeel Razi[2], Guillaume Flandin[1], Jean Daunizeau[3], Oliver J. Hulme[4,5], Alexander J. Billig[6], Vladimir Litvak[1], Cathy J. Price[1], Rosalyn J. Moran[7], Anthony Costello[8], Deenan Pillay[9], and Christian Lambert[1]

[1]*The Wellcome Centre for Human Neuroimaging, University College London, UK*
[2]*Turner Institute for Brain and Mental Health & Monash Biomedical Imaging, Monash University, Clayton, Australia*
[3]*Institut du Cerveau et de la Moelle épinière, INSERM UMRS 1127, Paris, France*
[4]*Danish Research Centre for Magnetic Resonance, Centre for Functional and Diagnostic Imaging and Research, Copenhagen University Hospital Hvidovre, Kettegaard Allé 30, Hvidovre, Denmark.*
[5]*London Mathematical Laboratory, 8 Margravine Gardens, Hammersmith, UK*
[6]*UCL Ear Institute, University College London, UK*
[7]*Centre for Neuroimaging Science, Department of Neuroimaging, IoPPN, King's College London, UK*
[8]*UCL Institute for Global Health, Institute of Child Health, UK*
[9]*UCL Division of Infection and Immunity, University College London, UK*

**E-mails**: k.friston@ucl.ac.uk; thomas.parr.12@ucl.ac.uk; peter.zeidman@ucl.ac.uk; adeel.razi@monash.edu; g.flandin@ucl.ac.uk; jean.daunizeau@googlemail.com; oliverh@drcmr.dk; a.billig@ucl.ac.uk; v.litvak@ucl.ac.uk; c.j.price@ucl.ac.uk; rosalyn.moran@kcl.ac.uk; anthony.costello@ucl.ac.uk; d.pillay@ucl.ac.uk; christian.lambert@ucl.ac.uk

## Abstract

This technical report addresses a pressing issue in the trajectory of the coronavirus outbreak; namely, the rate at which effective immunity is lost following the first wave of the pandemic. This is a crucial epidemiological parameter that speaks to both the consequences of relaxing lockdown and the propensity for a second wave of infections. Using a dynamic causal model of reported cases and deaths from multiple countries, we evaluated the evidence models of progressively longer periods of immunity. The results speak to an effective population immunity of about three months that, under the model, defers any second wave for approximately six months in most countries. This may have implications for the window of opportunity for tracking and tracing, as well as for developing vaccination programmes, and other therapeutic interventions.

Technical report

# Contents

# Introduction

Over the past months, an alternative to standard epidemiological modelling has been considered in the form of dynamic causal modelling (Friston et al., 2020a). This approach inherits from statistical physics and variational procedures in Bayesian modelling and machine learning (Dauwels, 2007; Feynman, 1972; Friston et al., 2007; MacKay, 1995; MacKay, 2003; Winn and Bishop, 2005). The validity of this approach has been partly established in a series of reports looking at the role of population immunity within an outbreak in a single region (Friston et al., 2020a), the effect of population fluxes between multiple regions (in the United States of America) (Friston et al., 2020b)and the genesis of rebounds following lockdown, in relation to a second wave of infections (Friston et al., 2020c). In brief, the conclusions of this kind of modelling are: (i) population immunity—inherited from the initial phases of the pandemic—plays a key role in nuancing its subsequent progression (ii) in the context of population exchange between regional outbreaks, social distancing and lockdown strategies based upon the local prevalence of infection reduce morbidity and mortality. Finally (iii), the mechanism that underwrites a second wave depends sensitively on the rate at which population immunity is lost following the first wave. This affords a window of opportunity within which track and trace protocols may delay or defer any second wave until it can be rendered innocuous through vaccination or clinical advances (Aleta A et al., 2020; Chinazzi et al., 2020; Hellewell et al., 2020; Keeling et al., 2020; Kissler et al., 2020; Kucharski et al., 2020; Simonsen et al., 2018; Streeck et al., 2020; Wu et al., 2020). In this report, we consider a key question: how long is this window—or, equivalently, what is the period of effective immunity inherited at the population level from the first wave (Kissler et al., 2020).

Dynamic causal modelling[1] can be characterised as a generalisation of state-space modelling based upon differential equations. This contrasts with advanced descriptive approaches that fit curves to timeseries data, without any explicit reference to the underlying dynamics: e.g., (Tsallis and Tirnakli, 2020). Dynamic causal modelling differs from conventional epidemiological modelling in that it uses mean field approximations and variational procedures to model the evolution of probability densities—in a way that is similar to quantum mechanics and statistical physics (Greenland, 2006). This contrasts with epidemiological modelling that uses stochastic realisations of epidemiological dynamics to approximate probability densities with sample densities (Kermack et al., 1997; Rhodes and Hollingsworth, 2009; Vineis and Kriebel, 2006; White et al., 2007). One advantage of variational procedures is that they are orders of magnitude more efficient; enabling model inversion or fitting within minutes (on a laptop) as opposed to hours or days on a supercomputer (Rhodes and Hollingsworth, 2009). More importantly, variational procedures provide an efficient way of assessing the quality of one model relative to another, in terms of model evidence (a.k.a., marginal likelihood) (Penny, 2012). This enables one to compare different models using Bayesian model comparison (a.k.a. structure learning) and use the best model for nowcasting, forecasting or, indeed, test competing hypotheses about viral transmission.

---

[1] http://www.scholarpedia.org/article/Dynamic_causal_modeling

We have used this technology to build epidemiological models of how data are generated—in terms of latent causes like the prevalence of infection—that embed conventional epidemiological models (e.g., SEIR models: *susceptible*, *exposed*, *infected*, *recovered*) in an extended state space. For example, dynamic causal modelling allows certain probability densities to be factorized. A key example of this is to model a joint distribution over states of infection and clinical manifestation. In other words, instead of assuming that there is a difference between being infected (I) and having recovered (R), one can accommodate the fact that it is possible to express symptoms without being infected: e.g., a secondary bacterial infection following interstitial pneumonia (Huang et al., 2020). Conversely, one can be infected without showing symptoms. Crucially, dynamic causal models can be extended to generate any kind of data at hand: for example, the number of positive tests. This requires careful consideration of how positive tests are generated, by modelling latent variables such as the bias towards testing people with or without infection or, indeed, the capacity for testing, which may itself be time-dependent. In short, everything that might matter—in terms of the latent (hidden) causes of the data—can be installed in the model, including social distancing, self-isolation and other processes that underwrite transmission. When all such latent causes are included, model comparison can then be performed to assess whether they are needed to explain the data. Here, we leverage the efficiency of dynamic causal modelling to evaluate the evidence for a series of models that are distinguished by the rate at which effective immunity to SARS-CoV-2 is lost. This provides a probability distribution over the rate of loss that determines when, or if, a second wave will ensue (Friston et al., 2020c; Kissler et al., 2020). In what follows, effective population immunity refers to the proportion of people who cannot contract or transmit the virus. This means that the loss of effective immunity can be mediated in several ways, e.g., through viral mutation or increasing the size of the susceptible population, through population fluxes or spread of the virus into new regions.

Details about the dynamic causal model can be found in the above technical reports (Friston et al., 2020a; Friston et al., 2020b; Friston et al., 2020c). Please see Figure 1 and Table 1 for a summary of its form and parameters. The model was fitted to new cases and deaths using data available from Johns Hopkins University [2]. The inversion and subsequent model comparison used standard variational (Laplace) procedures (Friston et al., 2007; Marreiros et al., 2009), as implemented in academic (open source) software[3]. The particular model used here has a degree of face validity. Formal Bayesian model comparison—with the closest conventional epidemiological models—speak to a higher model evidence (Moran et al., 2020), i.e., it provides a more accurate and parsimonious account of the data via optimising a (variational) bound on model evidence. Its predictive validity has been partly established. For example, it predicted death rates would peak on 10 April in the United Kingdom, with an initial relaxation of lockdown on 8 May 2020. In what follows, we use dynamic causal modelling to ask a simple but crucial question: how quickly will immunity to SARS-CoV-2 be lost at a population level?

---

[2] https://github.com/CSSEGISandData/COVID-19.

Before addressing this question, we reiterate that this paper is a technical report illustrating how questions of this sort can be answered using variational Bayes and dynamic causal modelling. It explicitly does not purport to provide definitive answers. In other words, as the models are improved through Bayesian model comparison—or as more data become available—the inferences and posterior predictions below will change. Although these inferences are described definitively, they are entirely conditional upon the model used in this analysis, and the data available at the time of writing (8 June 2020).

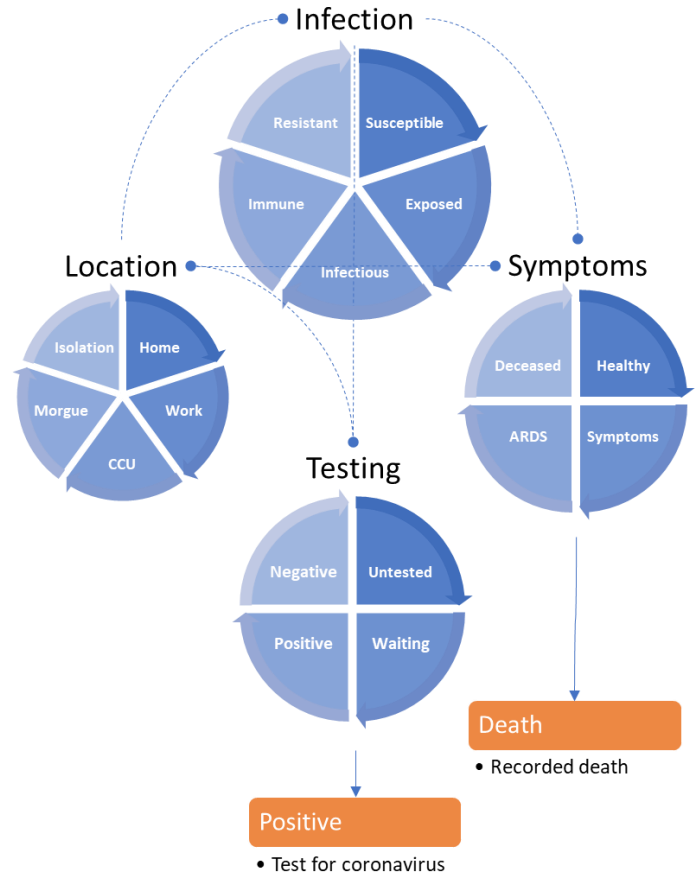

---

[3] https://www.fil.ion.ucl.ac.uk/spm/covid-19/

<div style="text-align:center">

FIGURE 1

</div>

**The LIST model**: This schematic summarises the LIST (*location*, *infection, symptom,* and *testing*) generative model used for the following analyses. This model is formally identical to that described in (Friston et al., 2020c). It includes a state (*isolation*) to model people who are self-isolating because they think they may be infectious (within their home or elsewhere). It also includes another (*resistant*) state to model individuals who are shielded or have pre-existing immunity, e.g., via cross-reactivity (Grifoni et al., 2020; Ng et al., 2020) or other protective host factors (Bunyavanich et al., 2020; Zheng et al., 2020). This absorbing state also plays the role of the *recovered* or *removed* states of SEIR models, namely, once entered, people stay in the state for the duration of the outbreak. One can leave any of the remaining states. For example, one only occupies the *deceased* state for a day and then moves to *healthy* (or *untested*) on the following day. Similarly, one only occupies the state of testing *positive* or *negative* for a day, and then moves to the *untested* state the following day. This ensures that the total population is conserved, e.g., deaths are offset by births into the susceptible state. Furthermore, it enables the occupancy of various states to be interpreted in terms of the rate of daily expression. The blue discs represent the four factors of the model, and the segments of these discs correspond to their states (i.e., compartments). The states within any factor are mutually exclusive, whereas the factors embody the factorial form of the compartmental model. In other words, every individual has to be in one of the states associated with the four factors or attributes. The orange boxes represent the observable outputs that are generated by this dynamic causal model, in this instance, daily reports of positive tests and deaths. The rate of transition between states—or the dwell time within any state—rests upon the model parameters that, in many instances, can be specified with fairly precise prior densities. These are listed in Table 1.

<div style="text-align:center">

TABLE 1

Parameters of the epidemic (LIST) model and priors, $N(\eta, C)$

(NB: prior means are for scale parameters $\theta = \exp(\vartheta)$)

</div>

| Number | Parameter | Mean | Variance | Description |
|---|---|---|---|---|
| 1 | $\theta_n$ | 4 | 1 | Number of initial cases |
| 2 | $\theta_r$ | 1/2 | 1/256 | Proportion of non-susceptible cases |
| 3 | $\theta_N$ | 8 | 1 | Effective population size (millions) |
| **Location** | | | | |
| 4 | $\theta_{out}$ | 1/3 | 1/256 | Probability of going out |
| 5 | $\theta_{sde}$ | 1/32 | 1/256 | Social distancing threshold |
| 6 | $\theta_{cap}$ | 16/100000 | 1/256 | CCU capacity threshold (per capita) |
| **Infection** | | | | |
| 7 | $\theta_{res}$ | 1/2 | 1/256 | Proportion of non-contagious cases |
| 8 | $\theta_{Rin}$ | 4 | 1/16 | Effective number of contacts: home |
| 9 | $\theta_{Rou}$ | 48 | 1/16 | Effective number of contacts: work |
| 10 | $\theta_{trn}$ | 1/3 | 1/16 | Transmission strength |

Technical report

| | | | | |
|---|---|---|---|---|
| 11 | $\theta_{inf} = \exp(-\frac{1}{\tau_{inf}})$ | $\tau_{inf} = 4$ | 1/256 | Infected period (days) |
| 12 | $\theta_{con} = \exp(-\frac{1}{\tau_{con}})$ | $\tau_{con} = 4$ | 1/256 | Infectious period (days) |
| 13 | $\theta_{imm} = \exp(-\frac{1}{\tau_{imm}})$ | $\tau_{imm} = 1:32$ | 1/512 | Period of immunity (months) |
| **Symptoms** | | | | |
| 14 | $1-\theta_{dev} = \exp(-\frac{1}{\tau_{inc}})$ | $\tau_{inc} = 16$ | 1/256 | Incubation period (days) |
| 15 | $\theta_{sev}$ | 1/32 | 1/256 | Probability of ARDS |
| 16 | $\theta_{sym} = \exp(-\frac{1}{\tau_{sym}})$ | $\tau_{sym} = 8$ | 1/256 | Symptomatic period (days) |
| 17 | $\theta_{rds} = \exp(-\frac{1}{\tau_{rds}})$ | $\tau_{rds} = 10$ | 1/256 | ARDS period (days) |
| 18 | $\theta_{fat}$ | 1/2 | 1/256 | ARDS fatality rate: CCU |
| 19 | $\theta_{sur}$ | 1/8 | 1/256 | ARDS fatality rate: home |
| **Testing** | | | | |
| 20 | $\theta_{ttt}$ | 1/10000 | 1 | Efficacy of tracking and tracing |
| 21 | $\theta_{lat}$ | 2 | 1 | Latency of sustained testing (months) |
| 22 | $\theta_{sus}$ | 4/10000 | 1/256 | Sustained testing |
| 23 | $\theta_{bas}$ | 4/10000 | 1/256 | Baseline testing |
| 24 | $\theta_{tes}$ | 8 | 1/16 | Selectivity of testing infected people |
| 25 | $\theta_{del} = \exp(-\frac{1}{\tau_{del}})$ | $\tau_{del} = 2$ | 1/256 | Delay in reporting test results (days) |

**Secondary sources** (Huang et al., 2020; Kissler et al., 2020; Mizumoto and Chowell, 2020; Russell et al., 2020; Verity et al., 2020; Wang et al., 2020). These prior expectations should be read as the effective rates and time constants as they manifest in a real-world setting. For example, a six-day period of contagion is shorter than the period that someone might be infectious (Wölfel et al., 2020)[4], on the (prior) assumption that they will self-isolate, when they realise they could be contagious. The priors for the non-susceptible and non-contagious proportion of the population are based upon clinical and serological studies reported over the past few weeks; e.g., (Ing et al., 2020; Stringhini et al., 2020). Please see the code base for a detailed explanation of the role of these parameters in transition probabilities among states. Although the (scale) parameters are implemented as probabilities or rates, they are estimated as log parameters, denoted by $\vartheta = \ln \theta$.

---

[4] Shedding of COVID-19 viral RNA from sputum can outlast the end of symptoms. Seroconversion occurs after 6-12 days but is not necessarily followed by a rapid decline of viral load. However, RNA shedding usually lasts longer than the shedding of infectious virus: many viruses produced are defective in some way, but still present RNA. As a rule of thumb 1/100 to 1/1000 virions produced are infective: van Kampen, J.J.A., van de Vijver, D.A.M.C., Fraaij, P.L.A., Haagmans, B.L., Lamers, M.M., Okba, N., van den Akker, J.P.C., Endeman, H., Gommers, D.A.M.P.J., Cornelissen, J.J., Hoek, R.A.S., van der Eerden, M.M., Hesselink, D.A., Metselaar, H.J., Verbon, A., de Steenwinkel, J.E.M., Aron, G.I., van Gorp, E.C.M., van Boheemen, S., Voermans, J.C., Boucher, C.A.B., Molenkamp, R., Koopmans, M.P.G., Geurtsvankessel, C., van der Eijk, A.A., 2020. Shedding of infectious virus in hospitalized patients with coronavirus disease-2019 (COVID-19): duration and key determinants. medRxiv, 2020.06.008.20125310.

# Results

The dynamic causal model above was fitted (i.e., inverted) using timeseries data from Johns Hopkins University[5] covering reported new cases and deaths from countries showing the highest cumulative number of deaths. The priors over the (25) model parameters can be found in Table 1. Crucially, this model inversion was repeated with different rates at which effective immunity is lost (i.e., the expected period of immunity following infection). These ranged from one month through to 32 months. This range was chosen to cover worst to best case scenarios. The worst-case scenario would correspond to a short-term period of immunity, less than that associated with the betacoronaviruses that cause the common cold: SARS-CoV-2 belongs to the betacoronavirus genus, which includes the SARS, MERS, and two other human coronaviruses, HCoV-OC43 and HCoV-HKU1 that cause the common cold (Kissler et al., 2020; Su et al., 2016). Immunity to HCoV-OC43 and HCoV-HKU1 appears to be lost over a few months. However, betacoronaviruses might induce immune responses against each another. For example, SARS can generate neutralizing antibodies against HCoV-OC43 that can endure for years, while HCoV-OC43 infection can generate cross-reactive antibodies against SARS (Chan et al., 2013). A period of 32 months corresponds to a level of effective immunity for close to three years, comparable to SARS-CoV-1.

The dynamic causal model used in this analysis accommodates heterogeneity of susceptibility and transmission at three levels, including a non-contagious proportion of the population that stands in for people who cannot transmit the virus. This inclusion speaks to the increasing appreciation of how heterogeneity in the population can have a fundamental effect on the epidemiological dynamics. This is variously described in terms of overdispersion, super spreaders, and amplification events (Endo et al., 2020; Lloyd-Smith et al., 2005; Paynter, 2016). In the current model, such heterogeneity was modelled in terms of three successive bipartitions (see Figure 2):

**Heterogeneity in exposure**: This was implicitly modelled in terms of an effective population size that is a subset of the total (census) population. The effective population is constituted by individuals who are in contact with contagious individuals. The remainder of the population are assumed to be geographically sequestered from a regional outbreak or are shielded from it. For example, if the population of the UK was 68 million, and the effective population was 39 million, then only 57% are considered to participate in the outbreak[6]. Of this effective population, a certain proportion are susceptible to infection:

**Heterogeneity in susceptibility**: This was modelled in terms of a portion of the effective population that are not susceptible to infection. For example, they may have pre-existing immunity, e.g., via cross-reactivity

---

[5] Available from https://github.com/CSSEGISandData/COVID-19. These timeseries were smoothed with a Gaussian kernel to suppress spurious fluctuations at the weekends.

[6] And of those 57%, some will be more exposed than others, conferring a further degree of heterogeneity, e.g., people working in care homes and hospitals, whose staff show a high seroprevalence: Houlihan, C., Vora, N., Byrne, T., Lewer, D., Heaney, J., Moore, D.A., Matthews, R., Adam, S., Enfield, L., Severn, A., McBride, A., Spyer, M.J., Beale, R., Cherepanov, P., Gaertner, K., Shahmanesh, M., Ng, K., Cornish, G., Walker, N., Michie, S., Manley, E., Lorencatto, F., Gilson, R., Gandhi, S., Gamblin, S., Kassiotis, G., McCoy, L., Swanton, C., Hayward, A., Nastouli, E., 2020. SARS-CoV-2 virus and antibodies in front-line Health Care Workers in an acute hospital in London: preliminary results from a longitudinal study. medRxiv, 2020.2006.2008.20120584..

(Grifoni et al., 2020; Ng et al., 2020) or particular host factors (Bunyavanich et al., 2020; Zheng et al., 2020) such as mucosal immunity (Seo et al., 2020). This non-susceptible proportion is assigned to the state of *resistance* at the start of the outbreak. Of the susceptible proportion of the effective population, a certain proportion can transmit the virus to others:

**Heterogeneity in transmission**: We modelled heterogeneity in transmission with a free parameter (with a prior of one half and a prior standard deviation of 1/16). This parameter corresponds to the proportion of susceptible people that cannot transmit the virus, i.e., those who move directly from a state of being *exposed* to a state of *resistance*, as opposed to moving from a state of being *infectious* to subsequent *immunity*. We associated this with a potentially mild illness—e.g., (Chau et al., 2020)—that does not entail seroconversion, e.g., recovery in terms of T-cell mediated responses (Grifoni et al., 2020; Zheng et al., 2020). Note that this construction conflates transmission with the probability of developing symptoms, in that being infectious means you can transmit the virus but also increases the period during which you could move from a *healthy* state to a *symptomatic* state.

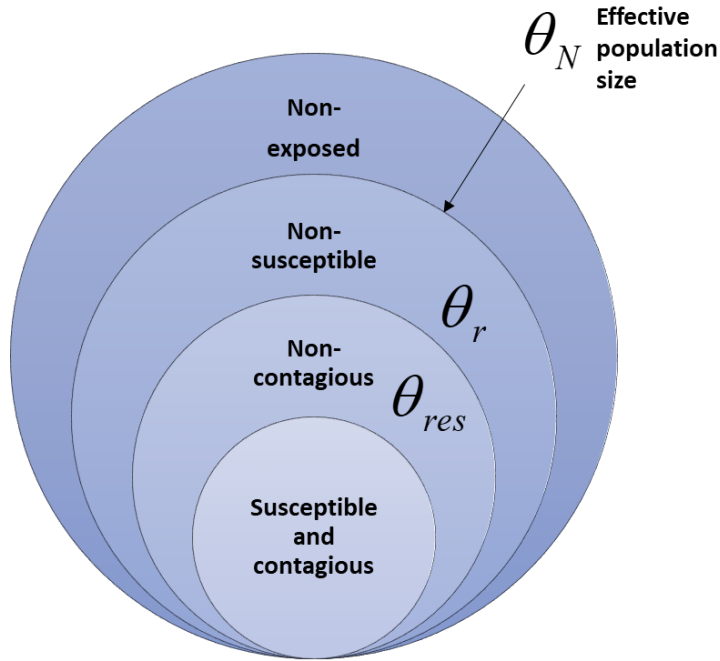

FIGURE 2

**Heterogeneity of exposure, susceptibility, and transmission**: Schematic illustrating the composition of a population in terms of people who are not exposed to contagious contact, not susceptible to contagion, susceptible but not contagious and, finally susceptible and contagious.

The *resistant* state therefore plays the role of an immune state for people who never become contagious, either because they are not susceptible to infection or become resistant after a mild illness. This model reconciles the apparent disparity between the relatively high morbidity/mortality rates and the relatively low seroprevalence observed empirically e.g., (Stringhini et al., 2020)[7]. Bayesian model comparison confirmed that there was very strong evidence (Kass and Raftery, 1995) for all three types of heterogeneity (portrayed as 'immunological dark matter' in the media); namely, an effective population that is a subset of the census population, a susceptible population that is a subset of the effective population and a contagious population that is a subset of the susceptible population (c.f., a super spreaders). In this model, only susceptible individuals who become contagious develop antibodies to SARS-CoV-2, typically around 8% of the total population.

Crucially, we did not impose any prior constraints on the effective population size[8]. In other words, we treated the data from each country as reflecting an outbreak in a population of unknown size that comprised a mixture of susceptible and non-susceptible individuals, where susceptible individuals comprised a mixture contagious and non-contagious individuals. In this way, we were able to model the self-evident dissociation between the total size of a population and the number of people affected in each country.

We specified a total of 32 models, each differing in their assumption about how long immunity would last, from 1 month to 32 months, in monthly increments. The log evidence for each of these 32 models was pooled over the 10 countries with the highest reported deaths (listed in Table 2). This evidence accumulation furnishes the marginal likelihood of each period of immunity (i.e., model) that—under uninformative priors over the period of immunity—corresponds to a posterior distribution, having marginalised out conditional uncertainty about all other parameters. Model inversion itself maximises the marginal likelihood that implicitly penalises overfitting, with respect to model complexity[9]. The resulting accuracy of the data fits are shown in Figure 3, in terms of cumulative death rates and new cases for the countries considered.

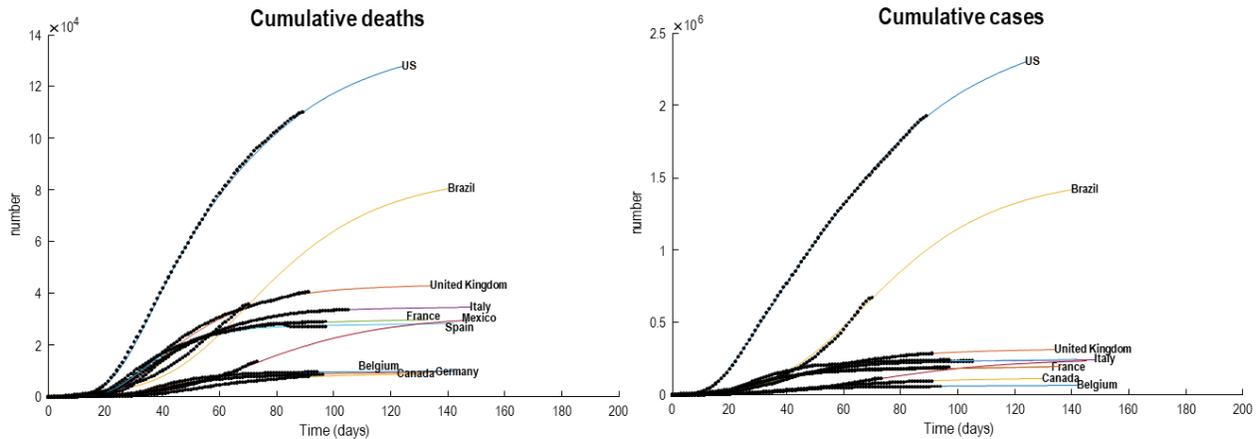

---

[7] See also: https://www.gov.uk/government/publications/national-covid-19-surveillance-reports/sero-surveillance-of-covid-19

[8] The susceptible population can be defined operationally as that proportion of the population that will eventually succumb to infection, with consequent immunity or resistance that may or may not be lost over time.

[9] Technically, the Kullback-Leibler divergence between posterior and the prior.

FIGURE 3

**Model accuracy**: this figure illustrates the accuracy of model inversion by plotting the empirical data for cumulative deaths (left panel) and cumulative new cases (right panel). The empirical data are shown as black dots overlaid on country specific predictions (coloured lines) based upon the latent states summarised in the subsequent figure. The trajectories have been shifted in time such that zero weeks corresponds to the time point at which the prevalence of the infection was estimated to be 0.1%

The accompanying distribution over the period of immunity is shown in Figure 4, suggesting that the expected period of immunity is about three months, with fairly precise 90% Bayesian credible intervals (less than the one month resolution of the model search). This does not mean that individuals will suddenly lose immunity after three months, rather that the effective population immunity will decline exponentially with a time constant of about three months. The 'effective' immunity refers to the fact that this characterisation of resilience is conditioned upon the model of aggregated or population dynamics. In other words, the effective population is behaving 'as if' its immunity is lost at this rate. There are many mechanisms that could contribute to this loss; for example, population fluxes could slowly increase the proportion of susceptible individuals (e.g., by relaxing lockdown); thereby diluting immunity acquired by the contagious proportion. Other viral and host factors (Beutler et al., 2007; Su et al., 2016) may clearly play a role (e.g., viral mutation or loss of antibodies)

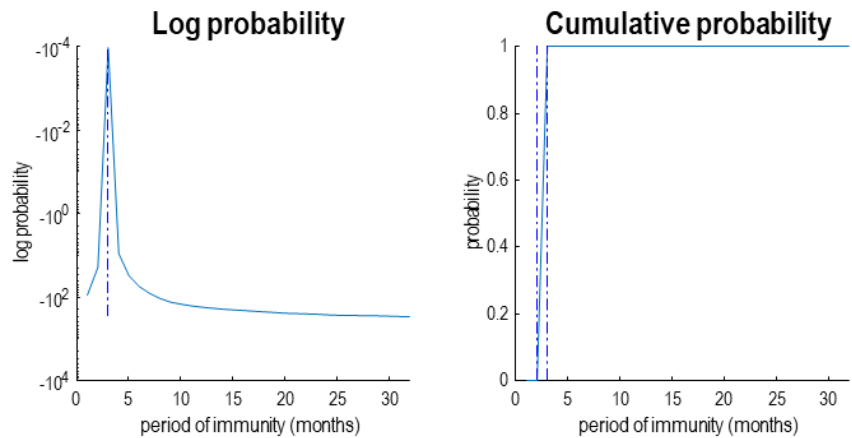

FIGURE 4

**Loss of immunity**: the left-hand panel reports the posterior distribution over the period of immunity based upon the marginal likelihood of a series of models that assume a particular prior expectation (with a precise prior covariance of 1/512). Here, the log posterior has been plotted on log scale. The form of the posterior over this key parameter reflects the fact that the trajectory of new cases and deaths contains sufficient information to make fairly precise inferences about the rate at which immunity is lost. The right-hand panel shows the same results in terms of a cumulative distribution. The broken lines correspond to 90% Bayesian credible intervals.

The rate at which immunity is lost is important because it constrains the onset of any putative second wave. Figure 5 illustrates this in terms of three scenarios for the effective population in the United Kingdom: first, a worst-case scenario with rapid loss of immunity (over a period of one month): a most likely scenario based upon the posterior expectation from Figure 4 (left panel) and, finally, a best-case scenario with a period of immunity lasting for years (32 months)[10]. We see that a very short period of immunity effectively merges the second wave into the first to produce a protracted time course of fatality rates. In effect, (a quasi) endemic equilibrium is obtained quickly as people lose immunity and become re-infected. With an immune period of three months, a second wave can be anticipated shortly after Christmas, in the New Year. With enduring immunity (here of 32 months) any second wave will be deferred for a year or more.

Interestingly, cumulative death rates appear to be higher with a three-month period of immunity, relative to a one-month period. This is because the predictions are posterior predictive densities, which are the most likely outcomes under the two periods of immunity. In other words, the best explanation for the data—under a one-month period of immunity—rests upon other model parameters that attenuate fatality rates, relative to a three-month period. Anecdotally, this kind of result suggests that we should be fairly confident about the loss of effective immunity in a month when the predictions under short (one and three months) and long term (32 months) scenarios diverge. One would be hoping to see death rates fall to negligible levels by October. If they persist at 20 deaths per day, then one might anticipate a second wave in January.

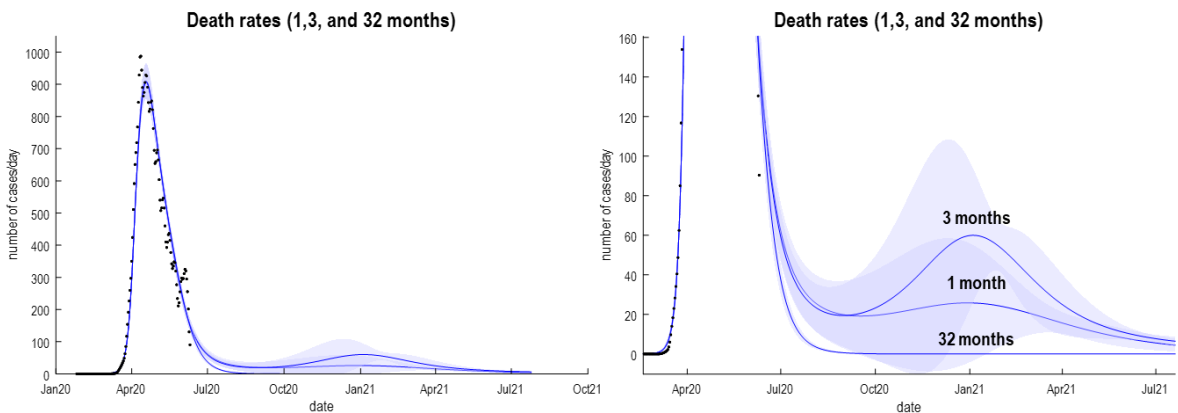

**FIGURE 5**

**Second waves:** This figure reports the expected death rates as a function of time for the effective United Kingdom population. The three trajectories (blue lines) and accompanying 90% Bayesian credible intervals (shaded areas) correspond to posterior predictions with a loss of immunity over 1, 3 and 32 months. These represent the smallest, most likely and longest period of immunity considered in the Bayesian model comparison (summarised in the previous figure). The black dots correspond to empirical data, after smoothing with a four-day Gaussian kernel. The right panel reproduces the data in the left panel with a focus on the second wave peaking in January of next year.

A common metric of viral spread is the effective reproduction ratio ($R_t$). This can be evaluated directly from the posterior expectations of latent states as a function of time. Figure 6 uses the same format as the previous figure to show the effective reproduction ratio for the United Kingdom. The initial fall in the effective reproduction ratio is subtended by lockdown in the first instance, followed by an acquisition of population immunity in the effective population. After reaching a minimum of about 0.7, the effective reproduction ratio slowly increases with loss of population immunity to peak in the late autumn, portending a second wave infections in January. Following this, the reproduction ratio remains largely below one and slowly drifts back to one after a year.

---

[10] Note that the best and worst scenarios are not determined by the credible intervals of the posterior distribution, but are the limits of the scenarios considered *a priori*.

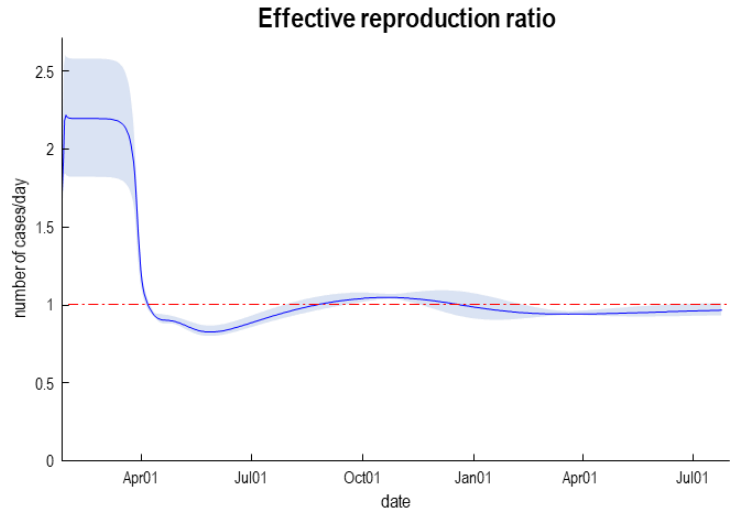

FIGURE 6

**Reproduction ratios**: this figure uses the same format as the previous figure; here, showing the predictions of the effective reproduction ratio for the United Kingdom. The initial fall in the reproduction rate is subtended by lockdown in the first instance, followed by an acquisition of population immunity in the effective population.

Figure 7 illustrates the underlying or latent causes of the predicted fatality rates over 18 months for the most likely (three-month) loss of immunity. These are the hidden states that we can infer from the modelling. In this model, the latent states are factorized into various *locations*, different states of *infection*, *symptom* expression and the states that underwrite the generation of *test* results. Please see figure legend for details. It is evident from these posterior predictions that the UK might expect a second wave in about eight months (around January 2021). This is important because there is a window of opportunity in the next few months during which nonpharmacological interventions—especially, tracking and tracing—will, in principle, be in a position to defer or delay the second wave indefinitely (or until an effective treatment or vaccination programme is in place). Please see (Friston et al., 2020c) for a more detailed treatment. Note that this model includes a latent state of *immunity* that peaks around 11% and then falls gently as immunity is lost (yellow line in the infection panel). In contrast, the *resistant* proportion (purple line) slowly accumulates people who *recover* from a mild illness and are *removed* from the susceptible proportion of the effective population.

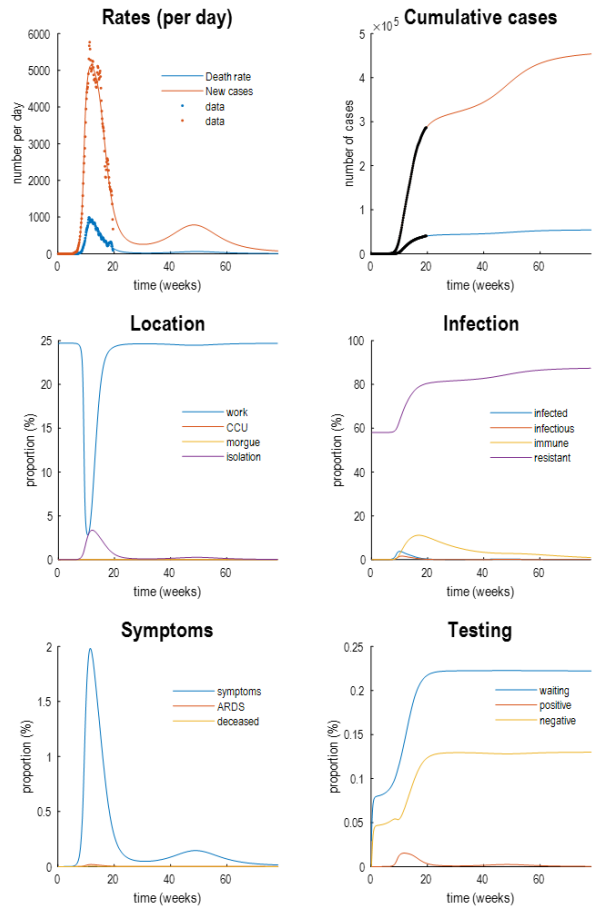

**FIGURE 7**

**Latent causes**: this figure reproduces the predictions of the second wave for the United Kingdom in Figure 4. Here, these outcomes are supplemented with the underlying latent causes or expected states in the lower four panels (the first state in each factor has been omitted for clarity: i.e., *home*, *susceptible*, *healthy,* and *untested*). These latent or expected states generate the observable outcomes in the upper two panels. The solid lines are colour-coded and correspond to the states of the four factors in Figure 1. For example, under the *location* factor, the probability of being found at work declines steeply from about 25% to 3% at the onset of the outbreak. At this time, the probability of isolating oneself rises to about 3% during the peak of the pandemic. After about six weeks, the implicit lockdown starts to relax and slowly tails off, with accompanying falls in morbidity (in terms of symptoms) and mortality (in terms of death rate). As population immunity (yellow line in the *infection* panel) declines, the prevalence of infection accelerates to generate a second wave that peaks at about 50 weeks. Note that the amplitude of the second wave is much smaller than the first.

The predictions above are generated from the parameters of a single country. However, these predictions conceal a large amount of between-country variability due to the non-linear relationship between the model parameters and trajectories of latent causes and states. Figure 8 shows the equivalent predictions of fatality rates for all 10 countries, under the most likely period of immunity (three months). Note that there is considerable variation in the onset of the second wave due to country-specific differences in the underlying epidemiological parameters. Table 2 summarises these differences in terms of the predicted dates of the first and second waves, respectively. For countries like the United Kingdom, this analysis suggests that one can anticipate a second wave in early 2021, which is later than the prediction for Germany, which might experience a second wave in October of this year (2020).

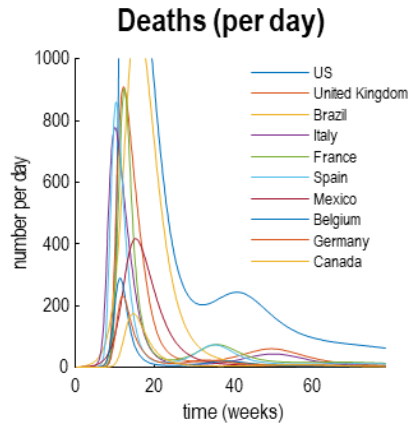

FIGURE 8

**Second waves over countries:** This figure illustrates the posterior predictive expectations of daily death rates for the countries considered at the time of writing (8 June 2020).

TABLE 2

List of countries and the dates of predicted fatality peaks of first and second waves.

| Country | First wave | Second wave |
|---|---|---|
| US | 24-Apr-2020 | 02-Nov-2020 |
| Brazil | 12-Jun-2020 | 27-Aug-2021 |
| UK | 18-Apr-2020 | 09-Jan-2021 |
| France | 16-Apr-2020 | 28-Sep-2020 |
| Spain | 10-Apr-2020 | 03-Oct-2020 |
| Italy | 06-Apr-2020 | 13-Jan-2021 |
| Mexico | 12-Jun-2020 | 29-Aug-2021 |
| Belgium | 19-Apr-2020 | 26-Oct-2020 |
| Germany | 24-Apr-2020 | 14-Sep-2020 |
| Canada | 14-May-2020 | 14-Feb-2021 |

This variation from country to country reflects differences in their epidemiological parameters. Table 3 summarises a few of these parameters and their variation. The first column lists the proportion of the effective population that are immune at the peak of population immunity. These range from 7 to 17% (4.5% to 5% of the total population), in line with current serological data[11]. The subsequent two columns make the point that the peak fatality rates at the second wave (based upon posterior predictions) are considerably smaller than the corresponding peak fatality rates at the first wave. For most countries, this second peak is in the order of tens of deaths per day, as opposed to hundreds.

The proportion of the susceptible population who cannot transmit the virus ranges from 47% to 61% (non-contagious column). This reflects heterogeneity in transmission. The corresponding heterogeneity of susceptibility is reflected in the proportion of the effective population that are not susceptible to infection (non-susceptible column). Finally, the difference between the effective and total population size reflects heterogeneity of exposure. In most instances, the effective population constitutes a large proportion of the total population (largest in Brazil, Spain, and Italy), with the exception of Canada, where the effective population is only four out of 38 million.

TABLE 3

List of countries and posterior estimates and population size. The non-contagious proportion is a percentage of the susceptible population, while the non-susceptible proportion is a percentage of the effective population. The numbers in brackets express the effective population as a percentage of the total (census) population.

| Country | Population immunity (percent) | First peak deaths (per day) | Second peak deaths (per day) | Non-contagious (percent) | Non-susceptible (percent) | Effective population (millions) | Total population (millions) |
|---|---|---|---|---|---|---|---|
| US | 15% | 2254 | 244 | 48% | 53% | 127 (38%) | 331 |
| Brazil | 7% | 1136 | 6 | 54% | 61% | 138 (65%) | 213 |
| UK | 11% | 988 | 60 | 57% | 58% | 33 (49%) | 68 |
| France | 17% | 964 | 74 | 49% | 53% | 19 (29%) | 65 |
| Spain | 11% | 863 | 73 | 54% | 66% | 33 (70%) | 47 |
| Italy | 10% | 819 | 44 | 61% | 59% | 37 (62%) | 60 |
| Mexico | 8% | 648 | 5 | 53% | 57% | 27 (21%) | 129 |
| Belgium | 16% | 331 | 20 | 52% | 52% | 5 (42%) | 12 |
| Germany | 15% | 266 | 24 | 47% | 59% | 13 (15%) | 84 |
| Canada | 13% | 174 | 8 | 60% | 47% | 4 (11%) | 38 |

---

[11] See also: https://www.gov.uk/government/publications/national-covid-19-surveillance-reports/sero-surveillance-of-covid-19

Technical report# Conclusion and limitations

There are clearly many limitations to the modelling here. These include modelling each outbreak as a point process and ignoring geospatial aspects and waves of infection (Chinazzi et al., 2020). Furthermore, we have assumed idealised dynamics that do not consider interactions with seasonal influenza or any other annual fluctuations (Kissler et al., 2020). As with all dynamic causal modelling studies, the conclusions based upon Bayesian model comparison and posterior inferences are limited to the models considered. Finally, the posterior predictions will change as more data becomes available. Having said this, it is interesting to note, irrespective of the modelling, that there is sufficient information—in the current epidemiological trajectories—to support fairly precise posterior beliefs about how quickly we will lose immunity.

Death rates in the United Kingdom over the next few weeks will be telling: if they can be suppressed to zero, then it is possible that the effective (population) immunity will be enduring, and we may elude a second wave. If, on the other hand, fatality rates continue above 20 a day, then according to the model presented here, it is likely we will see a slow increase in the reproduction rate and a second wave after Christmas. Note that the analyses in this report are predicated on a track and trace process whose efficacy is estimated based upon the data to date. As discussed in (Friston et al., 2020c) and elsewhere, any second wave could be deferred by introducing a more efficacious tracking and tracing protocol, even in the context of a relatively rapid loss of population immunity, such as the three month period estimated here. This deferment rests upon finding a substantial proportion of infected individuals before they can transmit the virus by identifying local outbreaks and clusters. On one view, this takes us out of the arena of ensemble dynamics and epidemiological modelling into the pragmatic considerations of an effective local surveillance and public health response.

## Software note

The figures in this report can be reproduced using annotated (MATLAB/Octave) code available as part of the free and open source academic software SPM (https://www.fil.ion.ucl.ac.uk/spm/), released under the terms of the GNU General Public License version 2 or later. The routines are called by a demonstration script that can be invoked by typing >> DEM_COVID_I at the MATLAB prompt. At the time of writing, these routines are undergoing software validation in our internal source version control system—that will be released in the next public release of SPM (and via GitHub at https://github.com/spm/). In the interim, please see https://www.fil.ion.ucl.ac.uk/spm/covid-19/.

The data used in this technical report are available for academic research purposes from the COVID-19 Data Repository by the Center for Systems Science and Engineering (CSSE) at Johns Hopkins University, hosted on GitHub at https://github.com/CSSEGISandData/COVID-19.

## Acknowledgements

The Wellcome Centre for Human Neuroimaging is supported by core funding from Wellcome [203147/Z/16/Z]. A.R. is funded by the Australian Research Council (Refs: DE170100128 and DP200100757). A.J.B. is supported by a Wellcome Trust grant WT091681MA. CL is supported by an MRC Clinician Scientist award (MR/R006504/1).

The authors declare no conflicts of interest.

Technical report